# Niche Modeling:
# Ecological Metaphors for
# Sustainable Software in Science

*Nicholas Weber, Andrea Thomer & Michael Twidale*[*]
*University of Illinois, Urbana-Champaign*
*{nmweber, thomer2, twidale@illinois.edu}*

## *Introduction*

An ecological metaphor is often used in discussing the interrelated and interdependent organization of software development environments. We believe that this is a rich and potentially useful way of describing complex systems of people and technologies that are involved in developing, using, sustaining and archiving software. Further, an appropriate metaphor can be a useful analytic lens for understanding any complex setting. By considering how well various components of the metaphor apply and fail to apply to aspects of a particular setting, we can deepen our understanding and uncover new questions to ask. In effect, a metaphor, when used in an appropriate way, can be an analytic creativity generator, provoking different questions and perspectives.

In the case of the ecological metaphor as used as a lens on scientific software development, many key ecological concepts remain underdefined and underutilized. For instance, might it be useful to think in terms of ecological niches? If so, what are the different niches commonly found in software ecologies, and how do different kinds of software fill them? What are the differences in trophic levels - or ecological efficiency - that make one niche of software development more successful than another?

This position paper is aimed at providing some history and provocations for the use of an ecological metaphor to describe software development environments. We do not claim that the ecological metaphor is the best or only way of looking at software - rather we want to ask if it can indeed be a productive and thought provoking one.

## *Ecological thinking*; **A brief history**

In 1948, ecologist G. Evelyn Hutchinson's paper "Circular causal systems in ecology" proposed that groups of organisms can be seen as a system of relationships with a circular causality (or feedback loop) that ensures the system or environment in which they live will survive over time (Taylor, 1988). As one group dominates or another withers and becomes extinct, the ecosystem adapts and new feedback loops are developed as a result. Hutchinson's proposal offers what we might now think of as a type of self-regulation where the sustainability of a large complex system is the result of its own organic maintenance.

---

[*] All authors contributed equally to this submission.

Applications of the feedback loop to non-biological systems became common nearly immediately after Hutchinson's paper was delivered in a 1945 symposium (Taylor, 1988). Contemporary examples of these not-strictly-biological ecologies include 'ecologies of the mind' (Bateson, 1980), information ecologies (Nardi and O'Day, 1999), ecologies of infrastructure (Star and Ruhleder, 1996), communication ecologies (Nevitt, 1982), and workplace ecologies (Walsh and Ungson, 1991).

The earliest use of "software ecology" that we could find was a paper from Barrington Nevitt in 1980. Expanding on his own previous attempts to define a communication ecology, Nevitt wrote "… information traveling at electric speeds has established a new software ecology in which the ground rules of the old hardware establishment break down, and cybernetic feedback can no longer keep pace with the action" (p. 225). Nevitt's use of "feedback" here echoes Hutchinson's original description of an ecological system, but he stops short of drawing any direct connection to a biological ecology, and it doesn't seem that he ever developed the concept further.

Later, McBride, Lambert, and Lane also used the metaphor, but mostly in relation to what we would now call human-computer interaction, "…execution of the algorithm's subroutines simulates the human, analytical process of representing abstract and concrete concepts. From an applied perspective, and particularly where human interactivity with machine-hosted information technology is concerned, the implications (software ecology, user-computer friendship, etc.) are difficult to ignore." (p. 247)

It seems that a 1981 IEEE tutorial from computer scientist Anthony Wasserman was the first to apply an ecological metaphor to the development of software. In describing changes in programming practices Wasserman wrote, "What is clear is that there is an ecology of software development environments. Changes in the environments usually have measurable direct and indirect effects." This explanation alone is salient, but Wasserman also goes on to propose that, "The challenge is to understand these effects and to create environments that have certain properties that promote cooperative and creative effort." (1981 p. 51).

As Wasserman's use of ecology hints, there are some problems with directly mapping Hutchinson's quantitatively derived theory of self-regulation in a biological sense, to a sociotechnical phenomenon like software development in an abiotic sense. Software of course does not develop itself; its evolution, spread, adoption, use and abandonment is always the product of human intention and agency.

This doesn't mean that a niche or ecology metaphor doesn't hold -- there are still external pressures that influence adoption, use and evolution -- but it does mean that instead of thinking about software ecologies as completely organic or naturally occurring we might instead consider software development akin to niche construction, restoration ecology, or ecological and environmental engineering in which systems are being intentionally altered (or in Wasserman's words, "created").

**Different ways the ecological metaphor may be used**
As with any metaphor, different people may apply ecology to the same setting, but for a variety of different purposes. In these cases there may be an aim to show that the metaphor helps account for much or all of what happens in a given setting – drawing direct, and clear relations between one domain and another.

In other cases, a metaphorical claim might hold less explanatory power; its use is for the sake of drawing some interesting parallels that can be helpful to consider, explore, push to extremes, and use to gain insight or to inspire further questioning and investigation. Our aim is around this latter approach. We don't want to claim that a software ecosystem is identical to a biological ecosystem and that all insights and methods can be transferred wholesale. Rather we want to ask, "What happens when you look at a software use-context as if it were an ecosystem? What are the issues that suddenly appear in focus? What do you start looking for as a result? What seems to match well, and what utterly fails to match?"  As Geoff Bowker wrote "if we are going to use the ecology metaphor… we should invest it with its full range of meaning" (2001).  We propose that investing software with a full  range of ecological meaning will allows us to:

- Focus on multiple pieces of software rather than on a single application
- Consider the interactions between, and interdependencies of software components
- Better explore the dynamics of introduction and extinction; or rather, the adoption of new software applications and the abandonment of others
- Witness ecologically contextualized evolution and co-evolution: how applications change over time in relation to one another
- Understand collaborative bricolage and tinkering – the ways in which people assemble and exploit resources to get work done more quickly and comprehensively.

We are particularly interested in how the concept of ecological niches might be used to shed light on software ecologies in science, and in turn, how this metaphor can be used to improve the design of systems and policies for their long-term sustainability.

**Niche Modeling**
The generally accepted definition of an ecological niche is, "the set of biotic and abiotic conditions in which a species is able to persist and maintain stable population sizes" (Hutchinson, 1957). In Hutchinson's account, there are two major classes of a niche:

1. Fundamental niche: the ideal range of conditions in which an organism could survive, barring interference from other taxa;
2. Realized niche: the actual range the organism is forced to occupy by pressure and interaction with other taxa.

For ecologists, estimating the "realized niche width" is a way of quantifying the space that an organism actually inhabits within any given ecosystem. We believe there are some intuitive applications of this suite of terms to scientific software engineering and studies of computer supported cooperative work (CSCW) more generally.

For instance, in their now famous CSCW paper *Steps towards an ecology of infrastructure*, Leigh Star and Karen Ruhleder discuss the failure of a scientific collaboratory to gain adoption amongst a group of loosely coordinated biology labs (1996). Using the ecological verbiage discussed above, this study can be reframed as a failure of the system to realize it's niche due to other selective pressures – the lack of IT support in some departments, the behavior and culture of information technologists, and an overall readiness for adoption amongst traditional biologists- all of which contributed to the collapse of the software niche.

Drawing inspiration from this work for scientific software, we might define a software niche as the set of technical requirements, organizational conditions and cultural mores that support its maintenance and use over time.  Fundamental niches might be defined as the range of conditions that scientific software is intended to occupy; whereas realized niches are their rates and modes of actual adoption and use. By better modeling available software niches – and their occupations -- we might better understand how to go about sustainable niche construction.

**Conclusion**
We believe that metaphor analysis, as described in this paper, can be a useful technique for understanding complex sociotechnical settings, and to inspire creative thinking about potential alternate designs or new lines of inquiry. In future work we plan to illustrate this approach using examples from two software platforms adopted widely in biology (R, and Matlab); in effect, showing how the modeling of different aspects of these software niches might help us better understand their use, maintenance and long-term sustainability.